# Wheeler-Feynman Absorbers on the Light Horizon

C. W. Lear[1]

[1]Dept. of Physics, Weber State University, Ogden, UT. 84403

Correspondence to:  charleslear@weber.edu

**Abstract:**  Early Wheeler-Feynman absorber theories invoke both retarded and advanced electromagnetic waves for photon emission and absorption in order to remove problems involving lack of radiative damping during electron acceleration.  Subsequent inquiries have suggested that only certain cosmologies would allow such a retarded-advanced wave mechanism to exist.  These include quasi-steady state cosmologies and exclude flat, expanding Friedman-type cosmologies.  Key to the exclusion process is a diminishing density of future absorbers in an ever-expanding universe.  Such absorbers would be expected to be real electromagnetically interacting particles.  However future virtual absorber sites, if they exist, would not be so diminished.  Such sites would be plentiful on the future light horizon, receding from the source at the speed of light.  The present treatment proposes that virtual absorption sites are present at every point in spacetime, and are characterized by the Fresnel-Kirchhoff diffraction integral.  On the future light horizon, they can remove electromagnetic energy from the local causal domain and provide advance wave signals as Wheeler-Feynman absorbers.

**1.0     Introduction**

In a classic 1945 paper Wheeler and Feynman [1] showed that if an electromagnetic emitter, such as an accelerated electron, sent out a wave front that was half retarded and half advanced the lack of radiative damping of the emitter acceleration was mitigated. Their article touched on the causal implications of the symmetry between emitter and absorber, and suggested the symmetry might be broken by the second law of thermodynamics.

Hoyle and Narlikar then, [2] and later in a book co-authored by Burbidge [3], treated the cosmological implications of the distribution of Wheeler-Feynman absorbers. They showed that most cosmologies could not support Wheeler-Feynman absorption because of the diminishing density of future absorbers in an expanding universe. They showed that a quasi-steady-state cosmology (QSSC) could supply future absorbers and could also break the symmetry of the emitter-absorber causality, which caused an ambiguity inconsistent with the second law of thermodynamics.

Hoyle, Narlikar and Burbidge have defended the QSSC with a variety of hypotheses, but the model is generally more complex than the current standard model cosmology, a flat expanding universe, and has not found wide acceptance.

In the present work we have formulated a model in which the diminishing density of real future absorbers in the current standard cosmology may be augmented by means of virtual Heisenberg resonant absorption. Virtual absorbers are ubiquitous throughout spacetime in the form of virtual electromagnetically charged pairs. They particularly appear on a "light horizon", when the local recession of the emitter is approaching the speed of light. On this light horizon, only the advance wave of the absorption-emission process will survive.

We will need to invoke the second law of thermodynamics as a symmetry breaking mechanism between past and future. This involves embedding the causal symmetry of the retarded-advanced wave field into the asymmetric propagation of thermodynamic probability into the future.

The next section will very briefly review the Wheeler-Feynman absorber theory and discuss the rationale for retrocausal action in advance waves. Following that we give a brief review of cosmological considerations. Then we will describe the virtual absorber model and apply some currently understood data to it.

**2.0    Wheeler-Feynman Absorbers**

When an electron emits radiation by changing its energy state in an atom, or by other means, in a semi-classical sense it undergoes acceleration. Such acceleration requires a radiative damping force. In an environment where there are no other moving charges there is no source for a damping field, and nothing to cause the electron to accelerate and radiate. Action-at-a-distance prevents any other neighboring charge to respond with a damping field with sufficient time and amplitude to satisfy conservation of energy and momentum.

Hoyle and Narlikar (2) described the situation as one with infinite self-action. For example, one could envision a very large unshielded electronic charge reacting to a very small self-induced field. The alternative, they said, was the acausality introduced by the quantum theory. Thus in a quantum field theory description of spontaneous emission, the need for a radiative damping field may be moot. Nevertheless, this is the situation addressed by Wheeler and Feynman [1] as they built upon the work of Schwarzchild and Fokker, and others, and it is worth following up here.

The Wheeler-Feynman approach showed that an advance wave from an absorber in the field was sufficient to provide the necessary radiative damping action precisely. In such case, the advanced and retarded fields must contribute in equal amplitude to the total field. In the nomenclature of Hoyle and Narlikar [2, pg 119], the field acting on an emitting charge (a) is the sum of advanced and retarded fields of all other charges (b), each of which may contain absorbers;

$$F_{total}^{(a)} = \sum_{b \neq a} \frac{1}{2}\left[F^{(b)ret} + F^{(b)adv}\right]$$

The implication here is that the absorbing charges (b) have a retrocausal action on the emitter (a). In particular such retrocausal action is symmetric under time reversal. A consequence of this symmetry is the action of the retarded wave on the absorber, supplying the radiative damping force needed to generate the advance wave.

In a thermodynamic environment that is asymmetric under time reversal, it is important to explore the implications of such retrocausal symmetry. Wheeler and Feynman addressed this at some length, and we intend to explore it again here.

In a previous discussion of Bell's theorem [4] we suggested that advance wave causality is a possible explanation for nonlocality and entanglement. Dynamic time reversal is a symmetry of field theory, wherein the time reversal operator T will change the sign of the dynamical time variable t without changing the state vector. If a state is symmetric under dynamical time reversal, it is reasonable that it must also be symmetric under causal time reversal. That is, the operator T not only changes the sign of t but it also changes the polarity of cause and effect.

The challenge to this symmetry is that retrocausation is inconsistent with the second law. If retrocausation emerges macroscopically,entails the consequence could be at least a limited violation of entropy increase within closed systems. Note that microscopic causality and microscopic dynamics are both time reversible, and we can claim they both obey the CPT invariance and Lorentzian invariance. Note, however, that microscopic causality and dynamics must translate to macroscopic causality and dynamics.

The difference is that the second law is based on a thermodynamic action principle of maximum probability evolution, whereas retrocausation is a symmetry principle of quantum mechanics. The two need not be incompatible. Symmetry is broken in thermodynamic systems but not in quantum systems. Quantum fields are governed by time reversible operators and wave mechanics. Thermodynamic action is probabilistic and results in an overwhelming preponderance of causal action in the direction of time asymmetry – a thermodynamic arrow of time. Living systems are thermodynamic systems and are propelled along this arrow of time insofar as sentient experience is concerned.

**3.0  The Cosmology**

Wheeler and Feynman [1] had proposed, at least tentatively, that a thermodynamic action principle is responsible for breaking the cosmological symmetry of causality. Hoyle and Narlikar [2], in contrast, declined to take this route and invoked a cosmological quasi-steady-state principle to maintain the needed symmetry for emission and absorption. If the second law, a maximum probability principle, results in a preponderance of forward causality in macroscopic dynamics, we can invoke this purely as a principle of cosmological thermodynamics, and not as a quantum principle.

The next serious challenge is the apparent inconsistency of the Wheeler-Feynman absorber theory with most currently accepted cosmological expansion models. They proposed, and Hoyle and Narlikar [2, pg 126] confirmed that the absorber theory requires perfect (future) absorption of retarded waves and imperfect (past) absorption of advanced waves in the cosmological time scale. Hoyle and Narlikar also find (ibid pg 139) that these absorption requirements are inconsistent with all the useful cosmologies.

They instead proposed quasi-steady-state cosmologies as a possible remedy, but these have suffered loss of credibility over time due to inconsistency with cosmological data [4]. They also showed, as did Wheeler and Feynman, that perfect future absorption is a requirement if absorber theory is to alleviate the dynamical deficiencies arising from photon emission by accelerated electrons. Hoyle and Narlikar (ibid pgs 125-126) showed the closed Friedmann cosmology as supporting both perfect past and future absorption, but that such overall perfect absorption leads to an ambiguity of cosmological causality of retarded and advanced waves. If thermodynamic asymmetry is rejected, the cosmology has no way of breaking the symmetry between past and future causality. Unwilling to accept this ambiguity, they discarded the closed Friedmann cosmology and sought solutions with imperfect past absorption.

Conceptually, perfect past absorption is easy to achieve because there is a much higher density of absorbers in the past, closer to the big bang. If a commensurate amount of future absorption is found, there is no need to invoke imperfect past absorption. Perfect future absorption in a closed universe is apparently achievable because radiation may recirculate throughout the enclosure until it is captured. Perfect future absorption may not occur in an open, expanding universe if the future absorber density is too small to capture the retarded radiation.

Our current cosmological data base strongly supports the existence of a cosmology that is open, flat and expanding. The Hoyle – Narlikar analysis as just discussed determined that, based on current models of the density of matter in the universe, there are not enough future absorbers to complete the needed absorption. But absorber theory, and consequently event theory, requires perfect future absorption. So if we are to proceed along these lines, we must look at other alternatives for future absorption. To develop these alternatives, let us explore some of the underlying cosmology.

Recent cosmological data indicate that the cosmological expansion is indeed accelerating (e.g. Perlmutter [5], Riess [6]). If so, the Hubble constant may be taken as increasing at some rate $\dot{H}_0$. (This is characterized in the literature as $q_0$.) We will deal with this accelerating cosmic expansion of a flat universe using the treatment of Carroll & Ostlie ([7] pg 1196, Eq. 29.131). They derive a linear scale factor function of time given by

$R(t) = 1/(1 + z) = \lambda_e / \lambda_o$

where t is the age of the universe, z is the cosmological redshift, and $\lambda_o$ and $\lambda_e$ are observed and emitted wavelengths. The functional form of R(t) is

$$R(t) = \left(\frac{\Omega_{m,0}}{\Omega_{\Lambda,0}}\right)^{1/3} \sinh^{2/3}\left(\frac{3}{2}H_0 t\sqrt{\Omega_{\Lambda,0}}\right). \qquad \text{Equ. 3.1}$$

The expression is derived from the Friedmann equation for an expanding universe, which in turn stems from Einstein's general relativity theorem. The expression deals with the expansion during the earlier mass era, which was dominated by gravitational attraction of mass, and the more recent $\Lambda$ era, which is dominated by the repulsive forces of dark energy. Appropriate modeling is put into the Friedmann Equation. Expansion during the earliest radiation dominated era, the relativistic era, is neglected. The $\Omega_{\_,o}$ density parameters are the ratios of the baryonic mass density to the critical mass density for the mass era and the $\Lambda$ era respectively. The critical density is that for which mass is expanding at exactly its escape velocity from collapse.

$H_0 = H(t_0)$ is the present value of the Hubble parameter which is a function of time. The expansion coefficient was derived for a domain of proper time t which remains uniform for all periods of the expansion, and is assumed to have a constant differential dt throughout the expansion.

### 4.0 The Light Horizons

As we look out into an expanding universe from our vantage point in the solar system, we are looking towards a rapidly receding sphere of influence moving away from us at the speed of light. Emitters arbitrarily close to this sphere, on our side of it, will be sending us photons which are red shifted arbitrarily close to oblivion, as determined by a Heisenberg uncertainty relation, $\lambda > d / 2\pi$, with d the distance of the emitter from the sphere of influence. Emitters on the other side of this sphere are receding from us faster than the speed of light, and so are their photons. We will never see them, and they are causally disconnected from us. We will refer to this expanding sphere as the past light horizon.

Photons of the cosmic microwave background (CMB) arriving in our domain of spacetime are coming from a distant past spherical shell called their surface of last scattering (Carroll & Ostlie [7], pg 1181). This nomenclature is due to the fact that most of them have scattered off of intervening matter in that shell.

Some of these photons will be scattered again in our domain, and will enter a new surface of present scattering. Some will leave our domain never again to be scattered. All will be joined by a relatively small amount of new radiant energy from present sources. The outward motion of energy from the surface of present scattering forms a spherical wave front, moving at the speed of light, which will be dimmed only by absorption. Absorption converts radiant energy to kinetic energy of massive charged particles. The surface of this outward moving wave front is chasing a spherical surface in space whose Hubble recession velocity gradually approaches the speed of light. We call this surface the future light horizon. When the surfaces coincide, both moving at light speed, any radiant energy becomes causally disconnected from our domain.

The future light horizon will appear to be moving outward, away from us in space, at light speed, just as is the past light horizon. The difference is that its approaching photons are moving away from us whereas those from the past light horizon are moving toward us. It has been shown by various authors (e.g. Hoyle & Narlikar [2]) that radiation from the surface of present scattering is not completely absorbed before reaching the future light horizon. We will show by conjecture that a process of virtual absorption of the radiation occurs at the future light horizon, resulting in the generation of advance waves returning to the source.

Photons from emitters in our present domain of spacetime are leaving a new past light horizon and entering a new surface of last scattering. They are moving outward toward a new future light horizon. Since the universe is expanding, there are already domains beyond our future light horizon which we cannot see. Their emitters and photons from those emitters are moving away from us faster than the speed of light, and we are causally disconnected from them. We specify that all events within past and future light horizons of a given domain are causally connected. Whether such connections are advanced, retarded, or both is the broader subject of this discourse.

Photons leaving the neighborhood of our solar system and bound for the future light horizon travel in an expanding sphere of diminishing intensity. The intensity is diluted principally by two factors; collisional and radiative damping. We will deal with each of these in turn. Each is affected by the cosmic expansion of spacetime.

Suppose the radius of a spherical surface surrounding local spacetime, $R = 1$, is given in meters as $\varpi$ (varpi). If $\varpi$ is taken as a co-moving coordinate, constant throughout time, (Carroll

& Ostlie, [7] pg 1148, Eq. 29.3), the radius of that surface at proper time t is given by $\varpi$ R(t). During a time increment dt the surface expands by an amount $\varpi$ dR.

We will examine the diminishing intensity of radiation within a cone extending from the source with arbitrary solid angle $\Omega$, which may be taken to be constant throughout cosmic expansion. We call this an absorption cone, because scattering and absorption by real particles occurs as radiation propagates through this cone. It originates at any point within a surface of last scattering and extends radially outward toward the future light horizon. Radiation entering the cone moves outward on a front that chases – and eventually catches – the future light horizon.

### 5.0 Intensity Loss by Area Expansion

Radiation intensity has a frequency spectrum governed by local emitters and which will assume a blackbody distribution in accordance with its expansion and time evolution into a cosmic microwave background. Intensity (I) is the ratio of power (P) and area (A) and each of these contributes to the time evolution of intensity loss. Power diminishes through collisional and radiative damping processes in the absorption cone. Area expands as the inverse square of the distance, and each of these is affected by the cosmic redshift expansion. We will examine the differential equation

$$\frac{dI}{dt} = \frac{\partial I}{\partial P}\Big|_A \frac{dP}{dt} + \frac{\partial I}{\partial A}\Big|_P \frac{dA}{dt},$$

which may be rewritten simply

$$\frac{d}{dt}\ln I = \frac{d}{dt}\ln P - \frac{d}{dt}\ln A.$$

The proper time t is the propagation time of the wave front in the absorption cone, beginning from zero at the surface of last scattering when the scattered energy enters the cone. We will attend first to the area expansion.

The area A satisfies

$$A = r^2 d\Omega$$

where $d\Omega$ is the constant solid angle of an elemental absorption cone. The distance r(t) is the proper distance from the vertex of the cone, called the emitter, to the wave front. The area increment dA has a factor $d\Omega$ which may or may not be an infinitesimal, and a factor $2\pi r dr$ where $dr = cdt$ is the increment of proper distance. A little algebra then shows that

$$\frac{d}{dt}\ln Ir^2 = power\ loss\ terms.$$

In the absence of scattering and absorption this is just the inverse square law.

Due to cosmic redshift expansion, the absorption cone wave front appears to diverge from its vertex faster than the speed of light, as viewed from the stationary frame of reference of the vertex. This is accounted for in using the proper distance for r(t) (Carroll & Ostlie, [7] pg 1203, Equ. 29.153):

$$r(t) = R(t) \int_{t_0}^{t} \frac{cdt'}{R(t')}$$

where $t_0$ is the present time, at the vertex of the absorption cone, when $R(t_0) = 1$, and R(t) is the cosmic expansion factor. In the local reference frame, r(t) accumulates at the speed of light, then expands by the factor R(t).

**6.0    Intensity Loss by Damping**

Collisional and radiative damping processes both contribute to the absorption of energy in the absorption cone. We refer to these as real absorption processes, because they are facilitated by real particles as opposed to virtual absorption processes that are facilitated by virtual particle pairs emergent under the constraints of the Heisenberg uncertainty principle. These will be discussed below. Real absorption processes convert radiant electromagnetic energy into kinetic energy of real particles.

Collisional and radiative damping processes occur over extended time periods, and both return advance waves to the radiant sources. As first explained by Wheeler and Feynman, the advance waves collect and collapse back to the source, beginning by gathering into a distant planar, semi-spherical, or other appropriately shaped wave front and collapsing to a sphere or other appropriate closed wave front back into the source. Thus, there is no need to maintain a unified integrated wave between emission and absorption. An important assumption in their work was that there were sufficient future absorbers to capture all the source energy, absorb it and return a complete advance wave front.

If we consider a fairly localized source at the surface of last scattering, or even at the surface of present scattering corresponding to the present age of the universe, it is useful to regard the divergence of radiation from the source as spherically symmetric and described by a Poynting vector such that the intensity is given in watts per square meter by:

**I = E x H**

Both **E** and **H** have propagation factors described by $\exp i(kr - \omega t)$ where $k = (\omega/c)(\eta - i\kappa)$ and $(\eta - i\kappa)$ is the complex index of refraction. The vectors **E** and **H** are perpendicular to each other, and are each perpendicular to the direction of wave travel. They are in phase with each other, and their vector product retains the same sign even as the individual vectors change sign. The result is that the wave frequency of the Poynting vector is doubled. The time averaged value of the magnitude of the Poynting vector is given by $I = \frac{1}{2} c\varepsilon_0 |\mathbf{E}|^2$, and the quantity $c\varepsilon_0 = 377\Omega$ is known as the impedance of free space.

The logarithmic differential equation for power propagation within an absorption cone then has the following form:

$$\frac{dP}{P} = 2i\omega \left(\frac{\eta}{c} dr - dt\right) - \frac{2\omega}{c} \kappa dr$$

The oscillatory term contains the real part of the index of refraction corresponding to wave velocity reduction through the plasma. The decay term is expressed as an imaginary part of the index of refraction. This consists of two subterms – one for collisional damping and the other for radiative damping. We will discuss each in turn.

### 6.1 Collisional Damping

A careful treatment of energy absorption by collisional damping would involve analysis of wave-particle interaction in the intergalactic plasma. The plasma is mostly a very rarefied ionized hydrogen. Particle-particle interactions are so rare we can treat the plasma as collisionless, and the dominant interaction is between electromagnetic waves and the lighter electrons. The mechanism, were we to choose it, would be collisionless Landau damping in which the wave particle interaction is strongest when the particle velocity is near the wave phase velocity. However this is beyond the scope of our need. We are concerned only with demonstrating that the damping processes are insufficient to remove all radiant energy from the absorption cones. We can get a rough analysis and a valid result by considering only the photon-electron interaction through Thomson scattering. This is the dominant mechanism of energy transfer. Thomson scattering diminishes the energy of the scattered photon and leaves the electron with extra kinetic energy on the average.

Collisional scattering is a thermodynamic process leading to increasing entropy and breaking the time symmetry between past and future. Wheeler & Feynman took the position that the asymmetry between retarded wave and advance wave propagation was due purely to thermodynamics and should not affect complete absorption of either retarded signals in the future or advance waves in the past. By contrast, Hoyle & Narlikar [2] explained that time symmetry was broken by cosmological expansion, and recourse to thermodynamics was not necessary. They chose to avoid the thermodynamic route, and to rely on the fundamental time symmetry of advanced and retarded waves. They looked to radiation damping as a means for energy transfer, as we will explain later.

The Thomson scattering process is well characterized with a cross section which is independent of energy but is dependent on scattering angle. The term in the complex index of refraction responsible for collisional damping may be directly replaced by this cross section, and we arrive at

$$\frac{d}{dt}\ln P = -c'n\sigma$$

where P is the scattered power, $c'$ is the speed of light in the scattering medium, n is the density of scattering particles and $\sigma$ is the scattering cross section. The derivation follows from considering the radiant energy entering a volume of constant cross section A and depth $c'\delta t$, and meeting a total scattering surface of $n\sigma$. Thus a proportion $n\sigma/A$ of the radiant energy is scattered out.

For the present value of the density of scattering centers in the universe, comprised mainly of ionized hydrogen in intergalactic space, we use the Wilson Microwave Anisotropy Probe (WMAP) value of 1 hydrogen atom per 4 m$^3$ given in Carroll and Ostlie [7]). For future redshift, as space expands volumetrically, this value is scaled down to

$n_H = .25/R^3(t)$ m$^{-3}$ .

Thomson scattering occurs off of free electrons. A good treatment of this mode is given in Jackson ([8], pg 679). Thomson scattering competes with inelastic Compton scattering of higher energy photons or Rayleigh scattering of longer wavelengths incident on atomic electrons, which we'll neglect. The total Thomson cross section is independent of wavelength with a constant value of $\sigma_T = 6.65\text{E-}29 \text{ m}^2$.

The dynamics of the scattering process are:

$h\nu = h\nu' + \frac{1}{2} mv^2$     for energy, with m the mass of the scattered electron;

$(h\nu' / c) \sin\theta = mv \sin\delta$     for transverse momentum with scattering angles $\theta$ and $\delta$;

$(h\nu/c) = (h\nu' / c) \cos\theta + mv \cos\delta$

The momentum balance gives

$\nu'/\nu = (\cos\theta + \cot\delta \sin\theta)^{-1}$

This ratio, confined between zero and one, is the fractional reduction in scattering cross section of photons in the ($\theta, \delta$) scattering channel. Integrated over the Thomson scattering cross section $d\sigma/d\Omega$ it yields the reduced total cross section for energy reduction. The result is about 12% of the total Thomson scattering cross section. Although a substantial reduction, this obviously leaves some unabsorbed radiation all the way to the light horizon.

*6.2    Radiative Damping*

When we introduced the idea of radiative damping above, it was as a means of providing action to accelerate an electron in an emitter. We sought to avoid an infinite self-action from decelerating an electron into a lower energy orbit, or put another way, an acausal emitter transition without any acceleration. We now note that absorption processes may also take place with radiative damping during which time the entire universe will react upon a charge to accelerate it into a higher energy state within a bound system.

The process is described in some detail in Hoyle & Narlikar [2]. They proposed radiative damping as a means of absorbing energy from primordial emissions. They concluded that the absorber density in the far future cosmic expansion was insufficient, and the absorption mechanism was not sufficiently robust to capture all radiation from primordial emissions. This led them to seek solutions in a closed universe. They concluded that the imaginary part of the index of refraction did not diverge to infinity, thus preventing complete absorption.

Hoyle & Narlikar (ibid) showed a simple, typical absorber equation of the form:

$$m\ddot{r} = eE + \frac{2e^2}{3}\dddot{r}$$

The third term is the effect of radiative damping on the absorber from all other oscillators in the universe. The equation shows the effect of such radiative damping along with the electric field **E** contributed by emitters. The effect applies to all absorbers of mass m and charge e. Due to the presence of second and third derivatives of absorber displacement, a Fourier transform of this equation will yield two independent equations from the real and imaginary parts separately. This places constraints on the imaginary part of the index of refraction as a function of the propagation parameters k and ω. These constraints may then be folded into the integral equation for field intensity, dependent on **E**, to give the development of the complex index of refraction over time and space.

**7.0    Virtual Oscillators and Secondary Wavelets**

Light is commonly described as an electromagnetic wave propagating in an extensive spacetime medium, with a well-defined spectrum of wave numbers and frequencies. There is, however, an alternative approach founded in potential theory. The relevant result is the Kirchhoff diffraction theory and the Kirchhoff diffraction integral.

The Kirchhoff theorem effectively treats every point in space as a virtual absorber-emitter of electromagnetic radiation. It is a refinement of the Huygens-Fresnel principle. In 1678, Christiaan Huygens proposed that every point that a luminous disturbance reaches becomes the source of a secondary spherical wave. The sum of such secondary waves, propagating forward in time, determines the form of the wave at any subsequent time. We refer to this as the retarded wave. Fresnel showed that the interference patterns from Huygens wavelets emerging from an aperture led to the diffraction of light.

The Kirchhoff diffraction approach describes the potential field at any point P, inside an arbitrary closed surface S, in terms of an arbitrary external source field F which is incident on S. The field internal to the surface is obtained by an integral, over the surface, of the incident external field and its derivatives. The surface is effectively covered with an array of infinitesimal oscillators excited and modulated by the external field. These are the source of what we will refer to as "secondary wavelets", some of which are Huygens wavelets.

The extended source field F will apply to either an electric or magnetic field. It is a solution of the Helmholtz wave equation

$$\nabla^2 F + k^2 F = 0.$$

The spatial field wave number, $k = 2\pi/\lambda$, may be a component of an extended spectrum. A Fourier analysis will separate the constituent frequencies in the time domain, and we will couch our equations without the frequency dependence, showing only wave number dependence.

It is instructive to examine parts of the derivation of the Kirchhoff diffraction integral, which begins with Green's theorem. The Green's theorem is applied to the external source potential function U and a secondary potential $U_0$ in a volume V surrounded by the surface S:

$$\iiint_V (F\nabla^2 F_0 - F_0 \nabla^2 F) dV = \oiint_S (F\nabla F_0 - F_0 \nabla F) \cdot d\vec{S}$$

The field of the secondary wavelet of unit amplitude and centered at P is designated

$$F_o = \frac{e^{ikr}}{r}$$

which also satisfies the Helmholtz wave equation. We take this to be a superposition of dipole oscillators, with directionality which will later be seen as dependent on its interaction with other oscillators within V and on S. The coordinate r is the length of a vector which conventionally points from P to S. Without loss of generality, we suppose that every point within V holds a real or virtual absorber-emitter. The site $F_0$ has one such oscillator.

Each of the two fields in the volume integral satisfy the Helmholtz equation with wave number k. The two terms in the integral are therefore identically equal, and cancel each other at every point in the volume. We are left with the surface integral on the right, which vanishes.

In the surface integrand, there is a singularity at point P, where r = 0. This point is surrounded by a small sphere to exclude it from the region surrounded by S. Thus, P is excluded from V. In the limit of small radius, integration over the surface area of the small sphere converges to a finite value, $4\pi F_P$. The fact that the singularity needs to be treated by exclusion implies that the oscillator there is a discrete, and this is an important part of our hypothesis. The completed integral now gives:

$$F_P = \frac{1}{2\pi} \oiint_S (F_0 \nabla F - F \nabla(F_0)) \cdot d\vec{S}$$

With a point source geometry, and the two restrictions r >> λ and ρ >> λ, the Fresnel-Kirchhoff approximation to the diffraction integral is appropriate. With S still arbitrary, we can gather a few factors into separate, convenient definitions and write:

$$F_P(\rho) = \frac{k}{2\pi i} F_s(\rho) \oiint_S (F_0(r) K(\rho, \bar{s}) \cdot d\vec{S}$$

$$F_s(\rho) = k^2 p \frac{e^{ik\rho}}{\rho}$$

Here, $F_s(\rho)$ is a superposition of fields of dipole sources normal to the rays ρ and, in the case of an extended source, normal to the surface vector of the source. We will refer to this type of dipole as a normal dipole, with constant amplitude $k^2p$, dependent upon interaction with neighboring oscillators. The obliquity factor,

$$K(\rho, \bar{S}) = \frac{1}{2}(\hat{n} \cdot \hat{r} - \hat{n} \cdot \hat{\rho}),$$

accounts for the relationship between P, at $\rho$, and the surface S. The vector $\bar{S}$ is a set of parameters defining the surface S, most generally in terms of continuous coordinates ($\rho$, $\theta$, $\phi$) defined on S. The unit vector $\hat{n}$ is normal inward everywhere on the surface S. The dot products of unit vectors are just the cosines of angles between those vectors.

The field $F_0$ is a dipole source, directionally modulated in amplitude by the source field and its derivatives, depending upon the point on S that terminates its r-vector. According to the Huygens-Fresnel formulation, the point P gives rise to an array of point oscillators on the surface S which are also modulated by the scalar external source and its derivatives.

With a point source geometry, and the two restrictions r >> $\lambda$ and $\rho$ >> $\lambda$, the Fresnel-Kirchhoff approximation to the diffraction integral is appropriate. With S still arbitrary, we can gather a few factors into separate, convenient definitions and write:

$$F_P(\rho) = \frac{k}{2\pi i} F_s(\rho) \oiint_S F_0(r) K(\rho, \bar{S}) \cdot d\vec{S}$$

$$F_s(\rho) = k^2 p \frac{e^{ik\rho}}{\rho}$$

The source field, $F_s(\rho)$, is written as that of a far-field normal dipole with moment p. It is effectively a point source. The obliquity factor,

$$K(\rho, \bar{S}) = \frac{1}{2}(\hat{n} \cdot \hat{r} - \hat{n} \cdot \hat{\rho}),$$

accounts for the relationship between P, at $\rho$, and the surface S. The vector $\bar{S}$ is a set of parameters defining the surface S, most generally in terms of continuous coordinates ($\rho$, $\theta$, $\phi$)s defined on S. The unit vector $\hat{n}$ is normal inward everywhere on the surface S. The dot products of unit vectors are just the cosines of angles between those vectors.

We now specialize the surface S as shown in Figure 1. Shown is a central cross-section of a right circular cone with interior containing V, bounded also by the inner and outer spheres centered at the cone vertex. The external potential is a point source at the cone vertex. The section of V is bounded in red, and the surfaces are so indicated. The coordinates for potential U are $\rho$ and $\theta$, centered at the point source. The point P is located at $\rho$, with $\theta = 0$. The surface coordinates are indicated with the cone half-angle $\theta_0$ and the sphere radii $\rho_{in}$ and $\rho_{out}$. The r-coordinate is shown at three alternative points on the surface S. The continuous coordinates $\theta$ and $\rho_{cone}$ serve to map points along the spherical and conical surfaces of S.

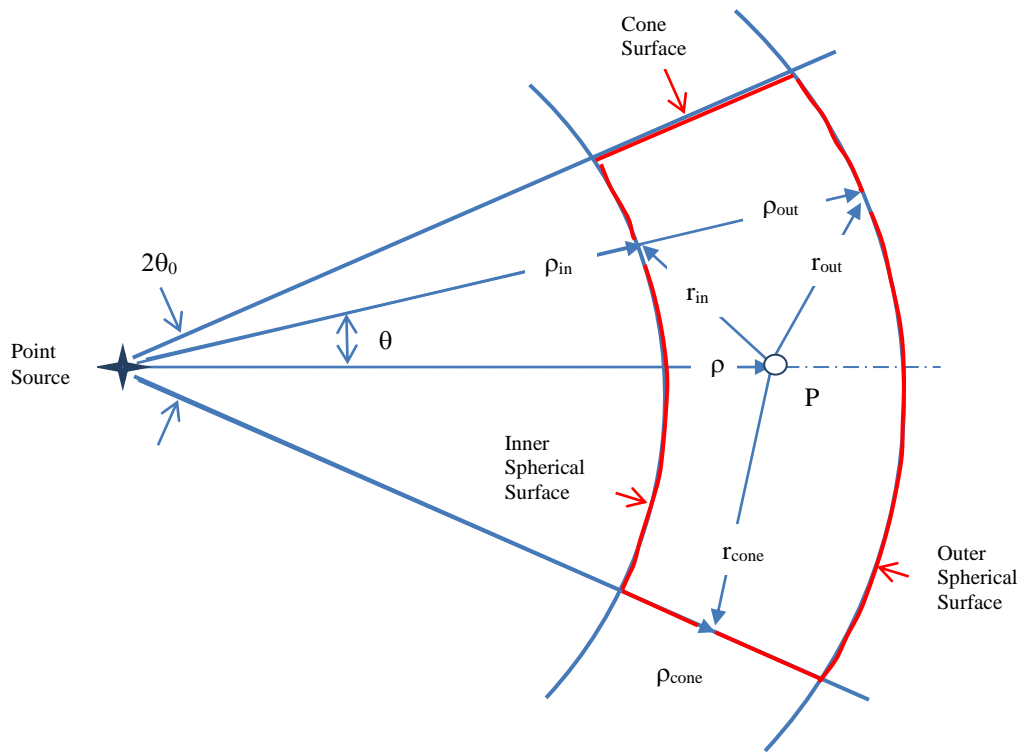

*Figure 1: Cross section of a Fresnel-Kirchhoff surface integral*

These coordinates, along with the discrete constants $\rho_{in}$, $\rho_{out}$ and $\theta_0$, make up the set $\bar{S}$ defining the conical surface.

Look at the obliquity factor on the surface of Figure 1. On the inner sphere, $\hat{n} \cdot \hat{\rho}_{in} = 1$ and the contribution to the obliquity factor is $\frac{1}{2}(1 + \hat{n} \cdot \hat{r}_{in})$. In the Fresnel-Kirchhoff surface integral, $\frac{1}{2}F_0(1 + \hat{n} \cdot \hat{r}_{in})$ is a dipole wave <u>from</u> a point oscillator at P modulated <u>to</u> a point (or annulus) at some surface radius $r_{in}$ and angle $\theta$. In the Huygens-Fresnel picture, $\frac{1}{2}F_0(1 + \hat{n} \cdot \hat{r}_{in})$ is a unit Huygens wavelet modulated <u>from</u> a surface oscillator at $r_{in}$ and $\theta$ <u>to</u> the point P. Can we reconcile the two?

On the inner sphere, the Huygens wavelets propagate in the forward direction, with amplitude diminishing to zero in the direction away from the surface and toward the source. On the outer sphere, the same situation appears, but the source ray and the surface normal are antiparallel, and $\hat{n} \cdot \hat{\rho}_{in} = -1$. The contribution to the obliquity factor is $\frac{1}{2}(-1 + \hat{n} \cdot \hat{r}_{out})$, showing unit Huygens wavelets propagating away from the source and out of V. Given the value of the surface infinitesimals $|dS|$ is the same on the inner and outer spheres, we can combine these two contributions, and the units, $\hat{n} \cdot \hat{\rho} = \pm 1$, cancel each other and the spherical, or normal, obliquity factor is

$$\frac{1}{2}(\hat{n} \cdot \hat{r}_{in} + \hat{n} \cdot \hat{r}_{out}).$$

Consider the situation on the surface of the cone. Here we have $\hat{n} \cdot \hat{\rho} = 0$ and the obliquity factor is $\hat{n} \cdot \hat{r}_{cone}$, which is a cosine function ranging from 1 to 0. The oscillators at P or S look like dipoles with axes oriented at angle $\theta_0$. The behavior of secondary wavelets on the cone is like absorption and re-emission from dipoles parallel to the source rays on the cone, and we will refer to these as parallel dipoles.

On an arbitrary surface, the ρ vector will be defined with components both normal and parallel to the incident ray. The normal component will generate Huygens wavelets and the parallel component will generate parallel dipole wavelets. They combine to recover the generalized obliquity factor given previously:

$$K(\rho, \bar{S}) = \frac{1}{2}\{\cos(\hat{n}, \hat{r}) + \cos(\hat{n}, \hat{\rho})\}$$

Given that the superposition depends on the three-fold relationship of the source potential and the oscillators on P and S, can we reconcile the three? Either the Fresnel-Kirchhoff surface is a convenient fiction, or the absorbers on the surface and at P are actualities. If they are actual, they may be real or virtual. If they are virtual, they lie under the shroud of the Heisenberg uncertainty principle. A way to actualize them is to hypothesize that the operative component at $F_0$ is an advance wave, and the superposition of normal and parallel dipole waves is a result of the radiative interaction between P and S. We conjecture that the amplitude ratios of the various waves is governed by the laws of conservation of electromagnetic momentum and energy.

This model of surface-to-volume interaction of oscillators in the Kirchhoff integral gives rise to a hypothetical physics of virtual or real oscillators arrayed throughout every point in space and time. Such oscillators are sources of retarded and advance waves upon excitation. Point source wavelets will superpose and merge to form classical wave fronts. We have worked with a point source, but an arbitrary external source field itself may be described as the sum of fields of an array of secondary wavelets. The point source $F_s$ then becomes a differential in a spectrum of wave numbers and frequencies emanating from points in a spacetime field.

Given a point P with field $F_p$, there is an arbitrarily large number of arbitrarily shaped surfaces surrounding it. We are led to question their interpretation and physical action. We propose the answer may lie in a multiple surface integral similar to what is done in the Feynman method for multiple path integrals over action. The field at P has a widely varying phase, depending on which surface contributes. There is one surface neighborhood with minimum variation, just as with the path variation giving rise to the principle of least action. All other surfaces will tend to contribute with destructive phase cancellations.

Not surprisingly, the extremal surface is a cone with vertex at the point source and directed to enclose the point P, with a cone angle approaching zero in the limit. The emergent wave is a single Huygens wavelet out of the source and exiting the cone at infinity. On the conic surface, $\hat{n} \cdot \hat{r} \to 0$, and there is no contribution. There is no contribution from the outer sphere, because $e^{ikr}/r \to 0$. The phase of the field at P is due entirely to the single Huygens wavelet, and there are no other interfering phase factors, destructive or otherwise.

To illustrate the multiple surface integral concept, consider a set of nested spheres centered on P, defined by vectors r with constant magnitude r on a sphere. The Fresnel integral shows that on a sphere $\hat{n} \cdot \hat{r} = 1$, and $\hat{n} \cdot \hat{\rho}$ / r is a slowly varying functional of r. The wave number k is large, so $e^{ikr}$ is a rapidly varying phase factor that tends to destructive cancellation at P over all the surfaces r. In like manner, integration over all other arbitrary surfaces will lend destructive interference from phase cancellations.

In the spirit and formalism of the multiple path integral (e.g. Feynman & Hibbs, [9] ) we can write an integral over multiple surfaces:

$$F_P = \oiint_S \frac{e^{ikr}}{r} K(r, \bar{s}) \mathcal{D}(S(r))$$

The integral is over multiple surfaces surrounding P, for which $\mathcal{D}$ is a differential operator selecting surface S(r), defined by vectors r of magnitude r and centered at P. The obliquity factor K is for a point source. The formalism $\mathcal{D}(x(t))$ is the same as used by Feynman & Hibbs, Equ. 2.25, to select a linear path x sequenced through time t. In the path integral, the phase factor $e^{ikr}$ replaces $e^{\frac{iS}{\hbar}}$ with S being the action integral over the time interval t. The path integral is a line integral over the coordinates x(t) rather than a surface integral over S(r).

The multiple surface integral clarifies how certain surfaces are excluded from the integration over $\mathcal{D}(S)$ because they are screened and have no phase contributions, leading to diffraction effects. We can see then how integration proceeds over extremal surfaces, neglecting the vast array of surfaces whose phase factors cancel to no contribution.

The absorbing centers are envisioned as pseudo-discrete points, meaning that absorbers cannot emerge arbitrarily close to each other but are separated by some small distance, perhaps of the order of a few classical electron radii. Quantization of spacetime could affect the determination of emergent sites, but that is not important for a semi-classical field. Candidates for such absorbers are electromagnetically charged pairs, of which the e+e- pair is the most energetically available, for example.

Emergent electron-positron pairs will absorb and re-emit dipole radiation. A simple classical model for a radiating pair gives two opposite charges orbiting around a center of mass. The force balance gives:

$m\omega^2 r = e^2 / 4\pi\varepsilon_0 r^2$,    $1/4\pi\varepsilon_0 = 8.99E9$ ( N-m$^2$ / C$^2$ )

The energy increment required to create such a pair includes the mass energy, the potential energy, and the kinetic energy of revolution:

$E = 2mc^2 - e^2 / 4\pi\varepsilon_0 r + \frac{1}{2} m\omega^2 r^2 = 2mc^2 - \frac{1}{2} e^2 / 4\pi\varepsilon_0 r$

Inspect these two equations for an e+e- mass of 1.02 MeV and a dipole frequency range from the mid IR to the upper UV, 200 – 3000 nm. The resulting radius of gyration, (r is not the Fresnel surface coordinate here), ranges over a fraction (0.11 to 0.68) of $r_0$, where $r_0$ is the classical electron radius 2.82E-15 m.  (Physically, $r_0$ is that separation where the potential energy of an electron exactly balances its mass energy.)  The energy increment is a deficit which must be borrowed from Heisenberg uncertainty at the expense of a very short lifetime of the pair.  The lifetimes are estimated from $\hbar/E$ and range from 33E-20 to 5.4E-20 seconds.  This is a very small fraction of the period of the dipole radiation, which is of the order of 1E-14 seconds.  If electron-positron pairs play a role in the propagation of Huygens wavelets they must do so with very short in-phase kicks to the dipole field, each of which adds a new wavelet with retarded and advanced potentials.  This may be expected, since the Aharonov Bohm effect tells us that the electromagnetic potential of a wave, including the phase, is physically real and will persist though the field disappears.  Emergent pairs are then expected to form dynamically in phase with the potential that forms them, and will result in field events that sustain the propagation of the wave.

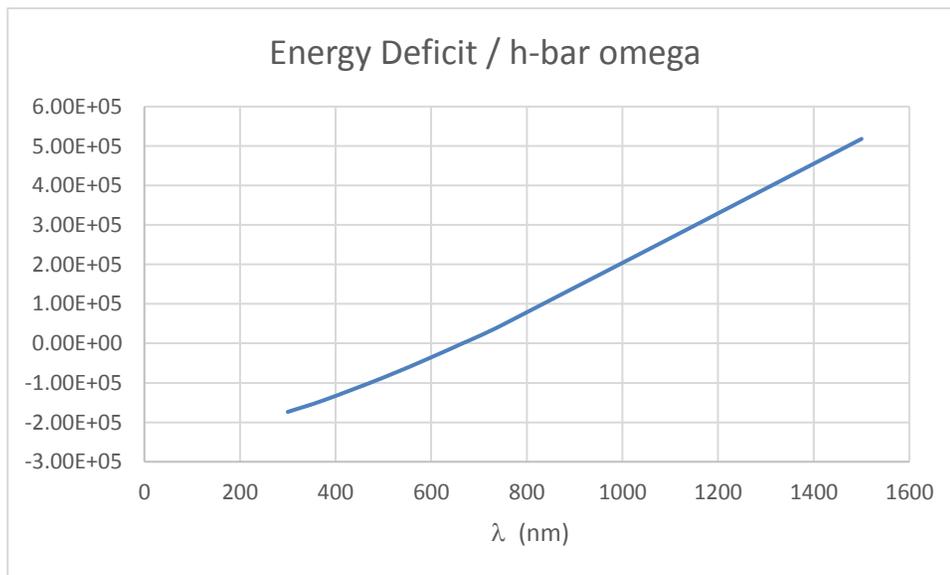

*Figure 2:  Giving the number of opportunities a virtual pair has to form during a period of the dipole wave of given wavelength*

The absolute value of the ratio $E/\hbar\omega$ is a ratio of times, and gives the number of times a virtual pair may potentially form, within Heisenberg uncertainty, during a period of the dipole wave. The ratio itself is plotted in Figure 2 as a function of the wavelength range under inspection. The high and low ends of the range have a very large number of potential field events – a half million or so. The number of potential events changes rapidly with changing wavelength, but in the neighborhood of 665 nm the required energy deficit gets very small and a pair can conceivably bootstrap itself into existence to support a number of full cycles of the dipole wave. This is not very important, since the vast majority of the domain of wavelength needs a very large number of potential field events to sustain it. Although such a large number of events is available, they need not all occur. The important thing is that they do occur often enough to sustain retarded and advanced field action on multiple surfaces.

The sign change of the ratio at 665 nm is simply due to the sign change of the energy deficit, which goes from borrowing energy from Heisenberg to get above the zero point, to yielding energy to Heisenberg to bring the electron and positron closer together with a more negative potential.

**8.0   Intensity Loss by Virtual Absorption**

A photon leaving a scattering site races toward its future light horizon with the speed of light, eventually catching up with it. The future light horizon begins as a spherical surface co-moving with the expansion of the universe with a radius increasing at somewhat less than the speed of light. According to the Hubble law, the recession velocity of the light horizon is accelerating, and eventually reaches, and then exceeds, the speed of light. At the point where the recession velocity equals the speed of light, the spherical wave front of outbound photons reaches it, and the radiation is absorbed. We will show that when the photon or wave front comes into coincidence with the light horizon, it is virtually absorbed on the horizon. When the photon or wave front comes into coincidence with the light horizon they are both traveling at light speed in the local reference frame.

As described by the Kirchhoff diffraction integral, Huygens-Fresnel wavelets are formed on the horizon surface. These combine to form an advance wave which propagates back toward the source. The combination also results in a retarded wave which propagates forward into the new causally disconnected spacetime domain. Huygens wavelets – or more generally, secondary wavelets – are always formed in pairs, advance and retarded. In each case, the wave front propagates away from the light horizon, toward the emitter in the incident domain and toward the absorber in the new domain – the absorber domain. The light horizon appears to propagate away from the emitter or absorber, in its reference frame, at the speed of light. There is a net transfer of momentum and energy across the light horizon.

Envision a spherical surface of present scattering from which an outward propagating wave front emerges. At $t_0$, the present age of the universe, on this surface, $R(t_0)$ equals one. It is enclosed by another concentric spherical surface of radius $\varpi$ which will eventually become the future light horizon. The radius $\varpi$ is called a co-moving coordinate (Carroll, pg 1148). It stays constant while the surface expands to a radius of $R(t)\varpi$. The expansion occurs over our local observer's time domain t, which has uniform time increments dt. At a future time $t_h$, photons from the outward propagating wave front which have not been scattered out will catch up to the future light horizon. The distance to the light horizon at any prior time, expressed as a proper distance in the Earth centered reference frame, may be written

$$L(t) = R(t) \varpi \qquad \text{Equ 8.1}$$

where the constant ϖ is the distance to a co-moving sphere, expanding with spacetime, which is to become the future light horizon. As such, the constant ϖ is also the initial value of the future light horizon at time t = 0, when R = 1. At the time $t_h$ the future light horizon is receding from the source at the speed of light. The co-moving coordinate is constant, so if we evaluate this equation at time $t_h$ :

$$(dL/dt)_h = \varpi \, (dR/dt)_h = c \,.$$

The time-dependent Hubble parameter evaluated at any future time t is given by:

$$H(t) = \frac{1}{R(t)} \frac{dR(t)}{dt}$$

(Carroll & Ostlie, [7] pg 1149 Eq. 29.8) This facilitates the solution of the co-moving coordinate,

$$\varpi = c \,/\, (dR/dt)_h = c \,/\, H(t_h) \, R(t_h) \,, \qquad \text{Equ 8.2}$$

and a solution for the position of the light horizon at the time of the critical recession velocity c:

$$L_h = c \,/\, H(t_h) \,. \qquad \text{Equ 8.3}$$

Also at this time the outward propagating wave front catches up with it. The distance traveled by the wave front in the expanding time domain is the proper distance, satisfying the condition (Carroll & Ostlie [7] pg 1203, Equ. 29.153):

$$L(t) = R(t) \int_{t_0}^{t} \frac{c \, dt'}{R(t')} \qquad \text{Equ 8.4}$$

where $t_0$ is the present time, at which $R(t_0) = 1$. When Equations 8.2 and 8.3 are set equal, the solution is $t = t_h$, the time for absorbed photons to reach their light horizon.

Using the Carroll-Ostlie approximation for the expansion coefficient and performing the differentiation of R(t) yields the inverse Hubble parameter,

$$\frac{R}{dR/dt} = \frac{1}{H_0\sqrt{\Omega_{\Lambda,0}}} \tanh\left(\frac{3}{2}H_0 t\sqrt{\Omega_{\Lambda,0}}\right).$$

This leads to a defining condition for the future light horizon,

$$R(t_h)\int_{t_0}^{t_h}\frac{cdt'}{R(t')} = \frac{c}{H_0\sqrt{\Omega_{\Lambda,0}}}\tanh\left(\frac{3}{2}H_0 t_h\sqrt{\Omega_{\Lambda,0}}\right) \qquad \text{Equ 8.5}$$

in which the left side is the distance traveled by the absorbed radiation and the right side is the distance to the absorbing light horizon. This equation may be solved numerically to obtain a solution for $t_h$. The results of such an analysis will be discussed in the next section.

The R / R-dot analysis shows rather clearly that the photon crosses the light horizon in a more-or-less discrete event. When the light horizon moves at less than the speed of light, photons are streaming past it. At the speed of light, it absorbs all its accompanying photons and accelerates on to speeds (as seen in our observer frame of reference) faster than the speed of light. At the light horizon, photons will continue to emit advance waves. Conservation of electromagnetic momentum dictates that the retarded wave continues to propagate forward into the next causal realm. but it will no longer be detectable in our frame of reference. We lose causal contact with those photons.

In a sense, photons at their light horizon discretely "wink out", if we could see them. We have no way of direct observation of advance waves. We can only watch the process by which they are emitted. This, of course, is done in the future of the emission. If we reject acausality of spontaneous emission, then the emission process itself is an indirect observation of the advance wave. Another indirect observation would be a verification of time reversal of cause and effect. If we ever develop a direct detection of advance waves, we could perhaps make direct observations of the dynamics of the future light horizon. This could provide a means of confirming or falsifying the hypothesis with a direct measurement.

## 9.0 Light Horizon Analysis

We want to show graphically how a spherical shell of radiant energy expands cosmically, always moving at the speed of light in its local frame, until it reaches its light horizon, which also expands cosmically from the emitter at the speed of light at the time of coincidence. When the radiant shell overtakes its light horizon, it is moving at the speed of light in the frame of reference wherein the light horizon is stationary. It is also cosmically expanding and moving at twice the speed of light in the emitter frame of reference.

To accomplish this we will need first to numerically solve Equation 8.5 for the proper time elapsed to reach the light horizon, denoted by $t_h$. Having done so, we will use Equation 8.2 to find the value of $\varpi$, varpi, the co-moving coordinate which is the initial radius of the future light horizon.

To solve for $t_h$, and subsequently to graph the motion of the radiant shell and the light horizon, we need parameters contingent to the equations of Sections 3. And 8.

Two recent values of the Hubble constant under review are the WMAP value of 73.8 ± 2.4 kilometers/sec/megaparsec and the European Space Agency Planck mission value of 67.3 ±1.2 kilometers/sec/megaparsec [3]. For our analysis, we take the present value of the Hubble constant as an average of 70.5 km/s/mpc. We will use the statistical uncertainty of the two treated as independent data points. A megaparsec has 3.089E19 kilometers, so we can write $H_0$ = 2.28E-18 $s^{-1}$. These and other parameters incident to the analysis of the light horizon are set down in the accompanying Table of Light Horizon Constants.

| TABLE OF LIGHT HORIZON CONSTANTS | | | |
|---|---|---|---|
| Symbol | Value | Source | Description |
| $H_0$ | 73.8 ± 2.4 km/sec/Mpc | WMAP | Current Hubble constant, equal to the recession velocity divided by distance from source |
| $H_0$ | 67.3 ± 1.2 km/sec/Mpc | ESA Planck | |
| $H_0$ | 70.5 ± 4.6 km/sec/Mpc | Average of WMAP & ESA Planck | |
| $H_0$ | 2.28E-18 sec$^{-1}$ | Average of WMAP & ESA Planck | |
| $t_H$ | 4.39E17 sec | 1 / $H_0$ | Current Hubble Time |
| $\Omega(\Lambda,0)$ | 0.73 ± 0.04 | WMAP | Baryonic to critical mass density ratio, during dark energy era |
| $\Omega(m,0)$ | 0.27 ± 0.04 | WMAP | Baryonic to critical mass density ratio, during gravitational attraction era |
| c | 2.998E8 m/s | International Consensus | Speed of light in a vacuum |
| Megaparsec | 3.089E22 meters | Conversion Factor | Arc seconds of earth orbit parallax |

Numerical solution of Equation 8.5 yields the proper time $t_h$ from emission for the radiant shell to reach its light horizon. The time is 7.82E17 seconds or 24.7 Gyr. The proper distance is 15.1E25 meters or 160E9 light years. These and other parameters are listed in the Table of Conditions at the Light Horizon, with time given in units of Hubble time and distance in megaparsecs.

Finally, with this data established, we can plot Equation 8.1 for the distance-time history of the light horizon and the integral Equation 8.4 for the distance-time history of the radiant shell. The radiant shell catches the light horizon as expected at the Hubble time of 1.781 and slopes upward at twice the velocity of the light horizon.

| CONDITIONS AT THE LIGHT HORIZON | | |
|---|---|---|
| Variable | Value | Description |
| $\tau$ | 1.781 $t_H$ | Proper time to light horizon |
| R | 2.057 | Cosmic expansion coefficient |
| dR/dt | 1.79 / $t_H$ | Rate of cosmic expansion |
| $H(\tau)$ | 1.99E-18 sec | Hubble constant at the light horizon |
| $\varpi$ | 2.37E3 Mpc | Initial radius of light horizon |
| $L_h$ | 4.88E3 Mpc | Proper distance to light horizon |
| $dL_h/dt$ | 3.00E8 m/s | Rate of light horizon recession |
| $dL'/dt$ | 6.00E8 m/s | Rate of recession of radiant shell |

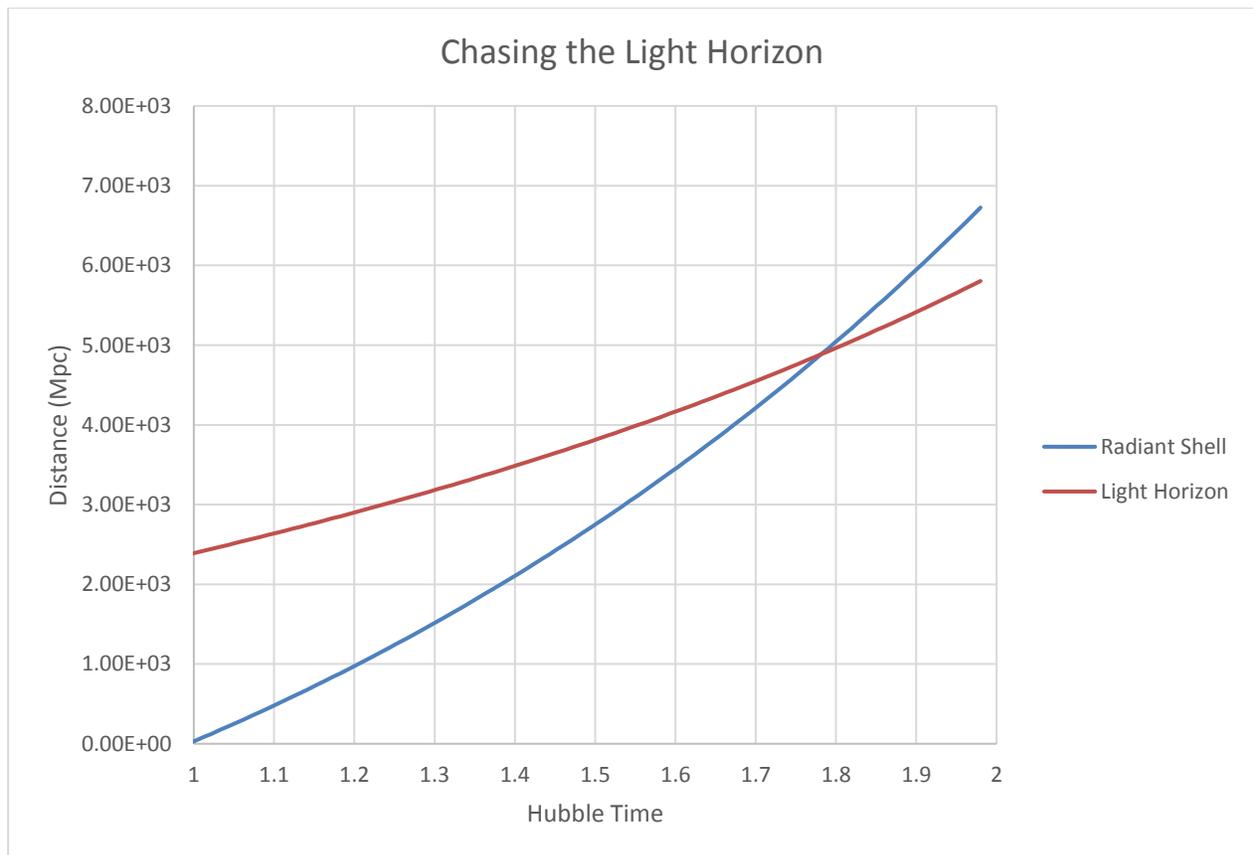

## 10.0   Discussion

The hypothesis shows a possible mechanism for future virtual absorption of photons emitted in the early and mid-term evolution of the universe which cannot find real absorbers.  In a flat, expanding universe the density of real future absorbers is too low to accommodate all early photons, including the copious efflux when the cosmos became transparent to electromagnetic radiation.

The absorption is virtual because the product of absorbed energy and interaction time is below the Heisenberg uncertainty limit.  Hypothetically, every site in spacetime is a discrete absorber-emitter of radiation due to the presence of a virtual electron-positron pair.  Every virtual absorption results in the re-emission of a wavelet with a retarded and advanced signal.  The exception is at the light horizon, where only the advance wave survives.  Summing over wavelets (Huygens-Fresnel wavelets) results in a field obeying the Helmholtz wave equation.

The light horizon is found by first finding the distance to a spherical surface in spacetime that is receding from the source of radiation at the speed of light, due to cosmic expansion of spacetime.  This distance is set equal to the distance traveled by a radiant shell traveling at the speed of light through the expanding spacetime.  The radiant shell overtakes and crosses the light horizon at light speed, leaving the realm of causal contact of the emitting sources.  A numerical solution of the equations yields a crossing at 1.79 Hubble times and a distance of $4.88 \times 10^3$ megaparsecs, within the accuracy of the data.  At or before this crossing, every photon is absorbed and re-emitted sending out both a retarded and advance wave.  Retarded waves leave the realm of causal contact, and advance waves return to the source via the same path taken by the retarded waves from the source.  Absorption takes place predominantly with real absorbers.  Residual radiation approaching the light horizon are finally absorbed by virtual electromagnetic pairs under the constraints of the Heisenberg uncertainty principle.

Future absorption of all emitted radiation allows the Wheeler-Feynman absorber theory to account for radiative damping needed for emission by accelerated electrons.  The absorber theory depends on advance wave causality in a fundamental way.

## Acknowledgments

The author gratefully acknowledges the help of Brad Carroll who inspired the idea of the light horizon. I would also like to thank Garrett Moddel for advice in publishing this work.

____

## I woReferences